\documentclass[12pt]{article}
\usepackage{a4}
\usepackage[dvips]{graphicx}
\usepackage{latexsym}
\usepackage{euscript}
\newcommand{\ifig}[1]{\includegraphics[height=8cm,width=14cm]{#1}}

\newcommand{\bc}{\begin{center}}
\newcommand{\ec}{\end{center}}
\newcommand{\be}{\begin{equation}}
\newcommand{\ee}{\end{equation}}
\newcommand{\bea}{\begin{eqnarray}}
\newcommand{\eea}{\end{eqnarray}}
\newcommand{\ba}{\begin{eqnarray}}
\newcommand{\ea}{\end{eqnarray}}
\newcommand{\amu}{A_{\mu}}
\newcommand{\m}{_{\mu}}
\newcommand{\ACal}{\cal{A}}

\newcommand{\simge}{\ \lower-1.2pt\vbox{\hbox{\rlap{$>$}\lower5pt
\vbox{\hbox{$\sim$}}}}\ }

\begin{document}
\centerline{\large {\bf{ Preliminary Results with Lattice Covariant Gauge}}}
\vskip 1.0cm
\centerline{L. Giusti$^{(1)}$, M. L. Paciello$^{(2)}$, S. Petrarca$^{(2,3)}$,
B. Taglienti$^{(2)}$}
\vskip 5mm
\centerline{\small $^1$ Boston University - Department of Physics, 590 Commonwealth Avenue} 
\centerline{Boston MA 02215 USA.}
\centerline{\small $^2$INFN, Sezione di Roma 1,
P.le A. Moro 2, I-00185 Roma, Italy.}
\centerline{\small  $^3$ Dipartimento di Fisica, Universit\`a di Roma "La
Sapienza",}
\centerline{\small P.le A. Moro 2, I-00185 Roma, Italy.}
\vskip 15mm
In this poster we present a few preliminary results obtained using our 
  method to fix  generic covariant gauges on the 
lattice.
We have computed the gluon 
propagator  and we have found a 
sensitive dependence on the gauge parameter.

As proposed in \cite{giusti,pisa}, 
the generic covariant gauge can be defined by replacing the Landau gauge condition
with the following form:
\be\label{dinamic1}
\partial\m\amu^G(x)=\Lambda(x)\; + ({\rm periodic} \; {\rm boundary} \; {\rm conditions}) 
\ee
where $\Lambda(x)$ belongs to the Lie algebra of the
group.
The functional proposed in \cite{giusti} in order to fix non-perturbatively the
condition (\ref{dinamic1}) is
\be\label{eq:cov}
H[G]\equiv ||\partial_\mu A_\mu^G - \Lambda||^2 =
\int d^4x\mbox{\rm Tr}\left[(\partial_{\mu}A^G_{\mu}-\Lambda)  
(\partial_{\nu}A^G_{\nu}-\Lambda)\right]\; .  
\ee
In fact it has the property
\be\label{eq:deriva}
\frac{\delta H[G]}{\delta g}\propto \left[
D_{\nu}\partial_{\nu}(\partial_{\mu}A^G_{\mu}-\Lambda)\right]
\ee
with $ G(x)  = e^{i g^a(x) T^a}$, where $T^a$ are the Gell-Mann 
matrices. Eq.~(\ref{eq:deriva})
shows that $H[G]$ is stationary when the eq.~(\ref{dinamic1}) is satisfied. 
Spurious solutions, that can be generated by the minimization, do not
seem to influence our numerical results.  
On the lattice, in the generic covariant gauge eq. (\ref{eq:cov}),
the expectation value of a gauge-dependent operator ${\cal O}$ 
is obtained by
\be\label{eq:omedio}
\langle{\cal O}\rangle=\frac{1}{Z}\int d\Lambda
e^{-\frac{1}{2 \alpha}\int d^4x 
Tr(\Lambda^2)} 
\int dU {\cal O}({U^{G(\alpha)}})
e^{-\beta S(U)}
\ee
where $G{(\alpha )}$ is the gauge transformation that 
minimizes the discretized version of the functional (\ref{eq:cov}) and the
$\Lambda$s follow a gaussian distribution of width $\alpha$.
In fact, on the lattice, the correct
adjustment to the measure is built into the simulation recipe and there is no
need to compute the Faddeev-Popov determinant. 
Hence the numerical procedure implied by eq. (\ref{eq:omedio}) can be described as 
follows:
\begin{itemize}
\item  A gauge configuration
$\{U\}$ with periodic boundary conditions according to the gauge invariant 
weight $e^{-S_W(U)}$ is generated;
\item For each $\{U\}$ configuration random matrices
$\Lambda(x)$ belonging to the group algebra are extracted according to 
the gaussian weight of eq. (\ref{eq:omedio});
\item Given $\Lambda(x)$, a numerical algorithm minimizes a 
discretization of the functional
$H[G]$. That defines the lattice gauge fixing condition ;
\item The expectation value of the lattice gauge dependent operator 
is then given by the average over the configurations:
\end{itemize}
\be
\langle{\cal O}\rangle^{Latt}=\frac{1}{N}\sum_{\{conf\}}{\cal O}({U^G})\;\; .
\ee
This is the procedure we will use  to compute 
gauge dependent correlation functions in a generic covariant gauge.

Of course, in order to fix the gauge non-perturbatively on the lattice, it is
necessary to discretize the gauge fixing functional relevant for the gauge 
condition required. The freedom in the choice of the lattice definition of 
$A_\mu$, as discussed in ref. \cite{arte},
 can be used to build discretizations of the minimizing
functional which lead to an efficient gauge fixing algorithm. This was the
case for the standard algorithm of the Landau lattice 
gauge fixing. 
It is possible to take advantage 
of the freedom to choose the discretization of the gluon field to find a 
discretization of $H[G]$ (``driven discretization'') such that it takes only
a local linear dependence upon $G ( {\bar x})$. This aim can be
reached 
by choosing the discretization of each $H$ term in order
to guarantee the local linear dependence on $G({\bar  x})$ 
instead of following 
a particular definition of $A_\mu$. 
Using this idea $H[G]$ can be discretized in the following compact form 
\be
H_L(G)  =  \frac{1}{V a^4 g^2} Tr \sum_{x} J^G(x) 
J^{G \dagger }(x)\label{eq:HJJ}
\ee
where 
\be
N(x)=-8 I + \sum_{\nu} \left[ U_\nu^\dagger(x-\nu) + U_\nu(x) \right] ;\;
J(x)=N(x) -i a g \Lambda(x)\; .
\ee
It is easy to see that locally the functional transforms linearly in $G({\bar
x})$ and in the continuum limit it goes to the functional 
(\ref{eq:cov}). 
The functional (\ref{eq:HJJ}) is semidefinite positive and,
unlike the Landau case, it is not invariant 
under global gauge transformations.
The functional $H_L[G]$ can be minimized using the same iterative algorithm
adopted in the Landau gauge fixing. In order to follow the convergence of the
algorithm, two quantities can be monitored as a function of the number of
iteration steps: the functional $H_L[G]$ itself and
$$\theta_H = \frac{1}{V} \sum_x Tr [\Delta_H\Delta_H^\dagger] , \; \;
{\rm where} \; \;
\Delta_H(x)  =  \left[ X_H(x) - X_H^{\dagger}(x) \right]_{Traceless},$$ 
and 
\ba\label{eq:XH_1}
X_H(x) & = & \left[  \left( \sum_\mu U_\mu(x)J(x+\mu) + 
U^\dagger_\mu(x-\mu)J(x-\mu) \right) - 8 J(x) - 72 I\right.\nonumber\\
& + & 8 \left.( K(x) + K^{\dagger}(x) ) +
N(x) K^{\dagger}(x)  \right]\; . 
\ea
$\Delta_H$ is invariant under the transformations 
$\Lambda(x) \rightarrow \Lambda(x) + c$ like in the continuum. 
$\theta_H$ decreases (not strictly monotonically) reaching zero 
when $H_L[G]$ gets constant and it signals that the 
algorithm has converged.
\begin{figure}
\bc
\ifig{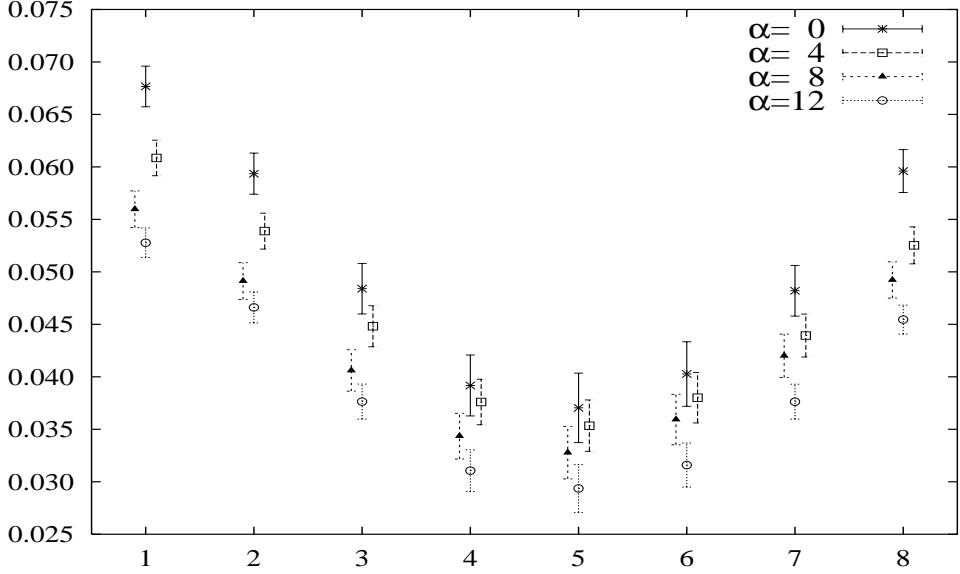}
\caption{\small{Correlations $\langle {\ACal}_i {\ACal}_i\rangle(t)$ for different
values of $\alpha$.}}
\label{fig:amu5}
\ec
\end{figure}
In this 
feasibility study we have generated 50, SU(3) thermalized 
configurations using the Wilson 
action with periodic boundary conditions 
at $\beta=6.0$ for a volume of $8^4$.
 Then for each value of $\alpha$ we have extracted a set of 
$\Lambda(x)$ with a gaussian distribution of width $\alpha$. 
We have monitored  
$H_L[G]$ and $\theta_H$ step by step and the chosen quality
is $\theta_H\le 10^{-10}$. 
Once the configuration have been gauge rotated we have computed
the following two point correlation function
\be
\langle {\ACal}_i {\ACal}_i\rangle(t) \equiv  \frac{1}{3 V^2}  
 \sum_{i=1,3}\sum_{{\bf x},{\bf y}} Tr \langle  A_i({\bf x},t)A_i({\bf y},0)\rangle .
\label{eq:AiAi}
\ee
In eq.~(\ref{eq:AiAi}) the standard definition 
is used for the 
gluon field.
Our result for the
 correlator $\langle {\ACal}_i {\ACal}_i\rangle(t)$ (\ref{eq:AiAi}), relevant
for the investigation of the QCD gluon sector, are shown in
 Fig.~\ref{fig:amu5}.
The relative statistical errors are comparable with the Landau gauge 
case with the same number of configurations.
Even with a small volume  and a small number 
of configurations, the gauge dependence of the gluon propagator 
is clearly shown.
The $\alpha$ dependence
of the gluon propagator shown in Fig.~{\ref{fig:amu5}} does not 
seem to be re-absorbed by an overall scaling factor.

 This plot shows the feasibility of our procedure to 
study the gauge dependence of physically interesting 
correlators.

\section*{Acknowledgements}
We thank the Center for Computational Science of Boston 
University where part of this computation has been done. This research was 
supported in part under DOE grant DE-FG02-91ER40676.
 S.P. thanks Valya Mitrjushkin and all the organizers of the workshop
"Lattice Fermions and Structure of the Vacuum" for the warm hospitality
and the excellent organization.

\end{document}